\documentclass[twocolumn,secnumarabic,amssymb, nobibnotes, aps, prl]{revtex4-2}
\usepackage{bm}
\usepackage{graphicx}
\usepackage[skip=8pt,font=scriptsize]{caption}

\begin{document}

\title{Resonance enhancement of the electromagnetic interaction between two charged particles in the bound state in the continuum}

\author{A.I. Agafonov}
\email{Agafonov\_AIV@nrcki.ru}
\affiliation{National Research Centre "Kurchatov Institute", Moscow 123182, Russia\\}
\affiliation{Moscow Aviation Institute (National Research University), Moscow, 125993, Russia\\} 
\date{\today}
\begin{abstract}
In bound states in the continuum (BIC) the mass of a composite particle is greater than the total mass of its constituents. Using the ladder Bethe-Salpeter equation, the BIC state in a system of two charged particles $a$ and $b$ with different masses ($m_{b}>>m_{a}$) is investigated. We demonstrate that for the positive binding energy, there are two momentum space regions in which the electromagnetic interaction between the particles is strongly enhanced, and the effective coupling constant turns out to be equal to $\alpha \sqrt{m_{b}/m_{a}}$, where $\alpha$ is the fine structure constant. The interaction resonance leads to confinement of the constituents in the BIC state with the positive binding energy which is of the order of the mass of the lighter particle from the pair.
\end{abstract}


\maketitle

\section{Introduction}
\par The coupling constant of the electromagnetic interaction is the fine structure constant, $\alpha=\frac{1}{137.04}$. Because of the relatively small value of the coupling constant, the mass of the composite particle $E$ in the normal bound states is slightly less than the mass of the constituent particles $\sum_{i}m_{i}$. Then the binding energy defined as $E-\sum_{i}m_{i}$, is negative, and, as a rule, is proportional to $\alpha^2$. Typical examples are the hydrogen and the positronium.  

\par However, there are unusual bound states which were discovered by von Neumann and Wigner in 1929 \cite{bib1} (see also \cite{bib2} with some extension of this work). In these states, which are called the bound states in the continuum (BIC), the composite particle mass is greater than the total mass of its constituents. Then the binding energy or, in other words, the mass defect is positive, $E-\sum_{i}m_{i}>0$, and can be much larger than that for the normal states.  

\par It so happened that the BIC phenomenon was forgotten for about 40 years \cite{bib3}. Thereafter, the BIC states have been found experimentally in condensed matter physics and optics (see  \cite{bib4,bib5,bib6,bib7,bib8} and references therein). These states are stable due to the confinement mechanisms that are individual for each case.  

\par In the non-relativistic mechanics, the BIC states have been investigated by using the Schrödinger Hamiltonian. As a rule, the spectral analysis of the real equation $H\psi=E\psi$ with $E>0$ was carried out \cite{bib3,bib9}. However, the  eigenvalues of the BIC states are in the continuous spectrum. Therefore, the study of the BIC states must be carried out with the use the Lippmann-Schwinger integral equation, in which an infinitesimal $\delta$ is added to the energy $E$. Then, the kernels of the integral equation and, respectively, the BIC wave functions become complex. As a result, the spectral analysis should be provided for a system of two coupled integral equations. 

\par As far as we know in quantum electrodynamics the BIC states have not been supposed and studied previously, and particle physics has been developed without the analysis of these states. The our idea is to apply this BIC phenomenon to some elementary particles. But the question arises: how can the electromagnetic interaction with the relatively small coupling constant leads to composite particles with large, positive values of the binding energy? This issue is discussed in the present paper.

\par The composite particle of two fermions is studied. Using the ladder Bethe-Salpeter equation, an integral equation for the BIC state of the two-particle system is derived. We demonstrate that the electromagnetic interaction between the particles is strongly enhanced when the momenta of the constituents are in the two regions in the momentum space. These momentum-space regions can be called the resonant regions because the interaction between particles becomes formally unlimited. Together with the correlations in particles motion, the resonance of the electromagnetic interaction leads to the confinement of these particles in the BIC state.

\par Natural units ($\hbar =c=1$) will be used throughout.

\section{Two particle system in the BIC state}
For the two-particle system, the Bethe-Salpeter equation can be written as:
\begin{eqnarray} 
\psi(1,2)&=&-i\int\int\int\int d\tau_{3}d\tau_{4}d\tau_{5}d\tau_{6}\nonumber\\& &K_{a}(1,3)K_{b}(2,4)G(3,4;5,6)\psi(5,6)~.
\label{eq:1}
\end{eqnarray}
In Eq.~(\ref{eq:1}) $d\tau_{i}=d{\bf r}_{i}dt_{i}$, $K_{a}$ and $K_{b}$ are the free propagators for the particles $a$ and $b$ that are assumed to be fermions, and $G(3,4;5,6)$ is the interaction function. In the ladder approximation the function is given by:
\begin{eqnarray}    
G^{(1)}(3,4;5,6)&=&-\alpha (1-{\bm \alpha}_{-}{\bm \alpha}_{+})\delta^{(4)}(3,5)\nonumber\\& &\delta^{(4)}(4,6)\delta_{+}(s_{34}^2)~,
\label{eq:2}
\end{eqnarray}
where $\alpha$ is the fine structure constant, and  $\delta_{+}(s_{56}^2)$ is the propagation function of the virtual photon. One of the function presentation is:
\begin{eqnarray} 
\delta_{+}(s_{34}^2)&=&\frac{1}{2\pi\vert {\bf r}_{3}-{\bf r}_{4} \vert}\lim_{\epsilon \to 0^{+}}
\Bigl[\int_{0}^{\infty}e^{-i\omega(t_{34}-r_{34})}d\omega\nonumber\\& &+\int_{0}^{\infty}e^{-i\omega(-t_{34}-r_{34})}d\omega \Bigr]~.
\label{eq:3}
\end{eqnarray}

\par For the problems of bound states, the free propagator for the particle $a$ takes the form:
\begin{eqnarray}    
K_{a}(1,3)&=&\sum_{\bf p}\frac{1}{2\varepsilon_{p}}\Bigl[ \Lambda_{a}^{+}e^{-i\varepsilon_{p}(t_{1}-t_{3})}+\Lambda_{a}^{-} e^{i\varepsilon_{p}(t_{1}-t_{3})}\Bigr]
\nonumber\\& &\theta(t_{1}-t_{3}) e^{i{\bf p}({\bf r}_{1}-{\bf r}_{3})},
\label{eq:4}
\end{eqnarray}
where
\begin{eqnarray}         
\Lambda^{\pm}_{a}({\bf p})&=&\varepsilon_{{\bf p}}\pm {\bm \alpha}_{a}{\bf p}\pm m_{a}\beta_{a},
\label{eq:5}
\end{eqnarray}
$m_{a}$ is its mass, ${\bf p}$ and $\varepsilon_{{\bf p}}=\sqrt{m_{a}^2+p^2}$ are the momentum and the energy of the particle, and the ${\bm \alpha}_{a}$ and $\beta_{a}$ matrices for the particle $a$ are taken in the standard representation.

\par Respectively, the free propagator of the particle $b$ is given by:
\begin{eqnarray}       
K_{b}(2,4)&=&\sum_{\bf q}\frac{1}{2\omega_{q}}	\Bigl[ \Lambda_{b}^{+}e^{-i\omega_{q}(t_{2}-t_{4})}+\Lambda_{b}^{-} e^{i\omega_{q}(t_{2}-t_{4})}\Bigr]
\nonumber\\& &\theta(t_{2}-t_{4}) e^{i{\bf q}({\bf r}_{2}-{\bf r}_{4})}
\label{eq:6}
\end{eqnarray}
Here 
\begin{eqnarray}       
\Lambda^{\pm}_{b}({\bf q})&=&\omega_{\bf q}\pm {\bm \alpha}_{b}{\bf q}\pm m_{b}\beta_{b},
\label{eq:7}
\end{eqnarray}
$m_{b}$ is the proton mass, ${\bf q}$ and $\omega_{{\bf q}}=\sqrt{m_{b}^2+q^2}$ are the momentum and the energy of the particle $b$. 

\par Let us assume that the binding energy $\cal E$ and the characteristic distance between particles are such that
\begin{eqnarray}       
{\cal E}r_{34}\leq \frac{1}{3}
\label{eq:8}
\end{eqnarray}
Neglecting this term in the exponents in  Eq.~(\ref{eq:1}), that means omitting the interaction retardation, we have:
\begin{eqnarray}       
\delta_{+}(s_{34}^2)&=&\frac{1}{\vert {\bf r}_{3}-{\bf r}_{4}\vert}\delta(t_{3}-t_{4}).
\label{eq:9}
\end{eqnarray}
As a result, in the momentum space, Eq.~(\ref{eq:1}) is reduced to the form:
\begin{eqnarray}       
\psi({\bf p},{\bf q};E)&=&-\frac{\alpha}{2\pi^2} K_{ab}^{(II)}({\bf p},{\bf q};E)
\nonumber\\& &\int \frac{d{\bf k}}{k^2}(1-{\bm \alpha}_{a}{\bm \alpha}_{b})\psi({\bf p}+{\bf k};{\bf q}-{\bf k}).
\label{eq:10}
\end{eqnarray}
Here the free two-particle propagator is:
\begin{eqnarray}       
K_{ab}&=&
\frac{1}{4\omega_{q}\varepsilon_{p}}
\Bigl[\frac{\Lambda_{a}^{+}\Lambda_{b}^{+}}{E-\varepsilon_{p}-\omega_{q}+i\delta}+\frac{\Lambda_{a}^{-}\Lambda_{b}^{+}}{E+\varepsilon_{p}-\omega_{q}+i\delta}
\nonumber\\& &\frac{\Lambda_{a}^{+}\Lambda_{b}^{-}}{E-\varepsilon_{p}+\omega_{q}+i\delta}+\frac{\Lambda_{a}^{-}\Lambda_{b}^{-}}{E+\varepsilon_{p}+\omega_{q}+i\delta}
\Bigr]
\label{eq:11}
\end{eqnarray}

\section{Resonance enhancement of the electromagnetic interaction}

\par Let $m_{b}>>m_{a}$. For calculation, we use $m_{b}= 1836 m_{a}$. In the BIC state, the total energy $E=m_{b}+m_{a}+{\cal E}$ with the positive binding energy, ${\cal E}>0$. 
Suppose that  ${\cal E}=\gamma m_{a}$ with $\gamma=1.531$. One can see that the imaginary parts of the last two terms in the square brackets in Eq.~(\ref{eq:11}) vanish. These terms are not 
essential for the formation of the BIC state and can be omitted. The first two terms in Eq.~(\ref{eq:11}) are fundamentally important. Using them, in Eq.~(\ref{eq:10}) we can introduce a 
function describing the effective interaction between the particles:    
\begin{eqnarray}       
\alpha_{int}&=&\frac{\alpha\Lambda_{a}^{+}\Lambda_{b}^{+}}{4\omega_{q}\varepsilon_{p}}
\Bigl[\frac{\cal P}{E-\varepsilon_{p}-\omega_{q}}-i\delta(E-\varepsilon_{p}-\omega_{q})\Bigr]+
\nonumber\\& &
\frac{\alpha\Lambda_{a}^{-}\Lambda_{b}^{+}}{4\omega_{q}\varepsilon_{p}}
\Bigl[\frac{\cal P}{E+\varepsilon_{p}-\omega_{q}}-i\delta(E+\varepsilon_{p}-\omega_{q})\Bigr]
\label{eq:12}
\end{eqnarray}

\par For normal bound states the energy eigenvalue is negative, ${\cal E}<0$. Then the $\delta$-functions in the right-hand side of Eq.~(\ref{eq:12}) vanish as well. Hence any enhancement of the interaction between particles does not occur in the convenient bound states. 

\par For the BIC state the energy eigenvalue is positive, ${\cal E}>0$. Hence $E-m_{b}\pm m_{a}>0$, and the term with the $\delta$-function is principally important. This function determines  
the two regions in the momentum space which are solutions of the equation:
\begin{eqnarray}         
(E-\omega_{\bf q})^2&=&\varepsilon_{\bf p}^2.
\label{eq:13}
\end{eqnarray}
Inside these regions the principal part 
\begin{eqnarray}           
\frac{\cal P}{(E-\omega_{q})^2-\varepsilon_{\bf p}^2}&=&0.  
\label{eq:14}
\end{eqnarray}

\begin{figure}[b]
\centering
\includegraphics[width=8cm]{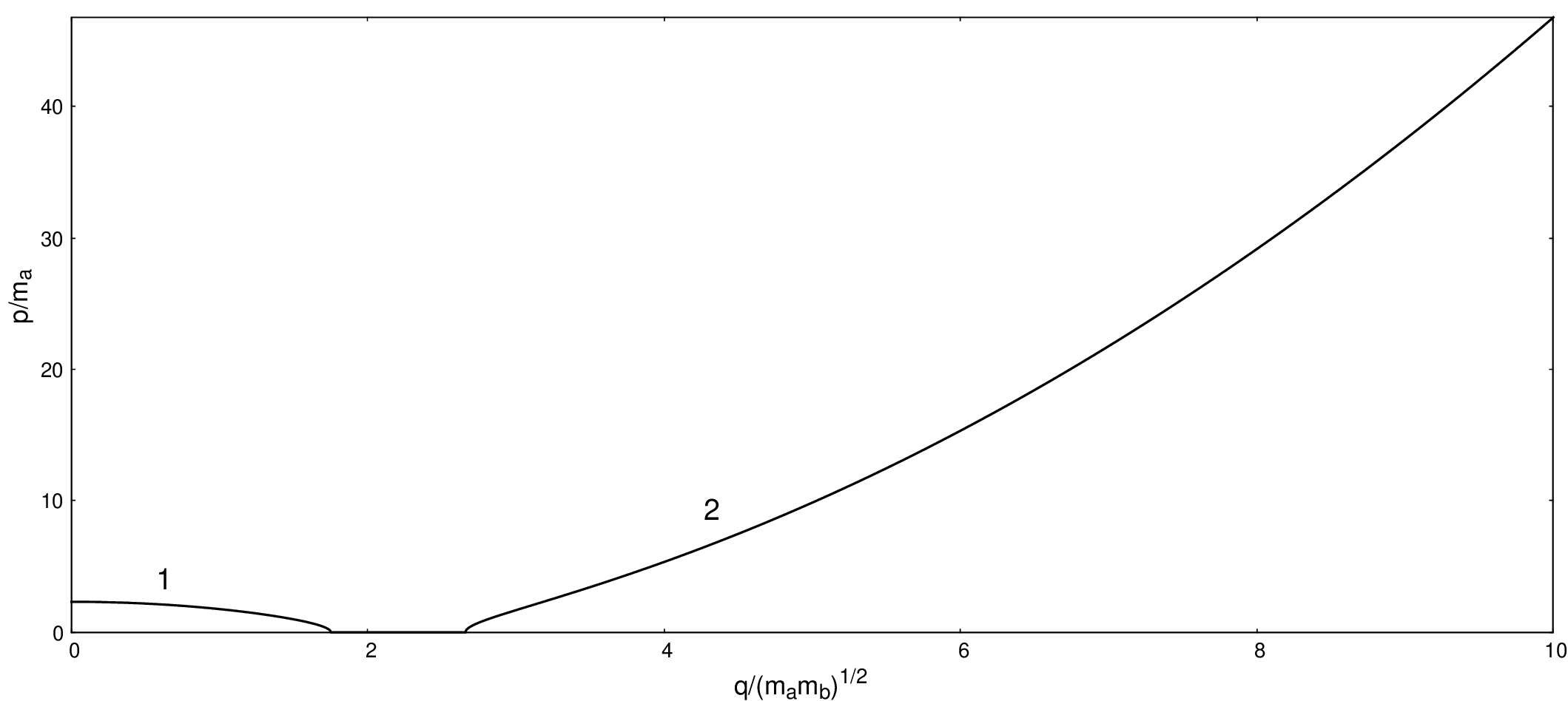}
\caption{The solution of Eq.~(\ref{eq:13}) for the energy $E=m_{b}+m_{a}+{\mathcal E}$ with ${\mathcal E}= \gamma m_{a}$. 
\label{f1}}
\end{figure}

\par The two lines in Fig. 1 correspond  the two regions in the momenta space. These lines are given the relation between the modules of the vectors ${\bf p}$ and ${\bf q}$. Their directions, ${\bf p}/p$ and ${\bf q}/q$ are also interrelated. In these regions, the motions of two particles are correlated with each other, and their interaction is sharply enhanced, becoming formally unlimited. Therefore, these regions can be called resonant ones.
 
\par The first resonant region is presented by curve 1 in Fig. 1. For the $a-$ particle, this region is limited from above by the moment $p=2.324m_{a}$, and for the the $b-$ particle the similar restriction is $q\leq 1.749\sqrt{m_{b}m_{a}}$. Here the energies of the both particles are positive. This curve is given by the first term in square brackets on the right side of 
Eq.~(\ref{eq:11}). However, this region is not important for the formation of the BIC state that is of interest to us. This is because the $a-$ particle momenta are relatively small. 
Respectively, it leads inevitably to large radius of the particle motion that is about of its Compton wavelength.

\par In this regard, the second resonant region represented by curve 2 in Fig. 1 are of undoubted interest. It is determined by the second term in square brackets on the right side of 
Eq.~(\ref{eq:11}). For this case, the $b$ particle energy,$\omega_{q}$, is above the lower boundary ($+m_{b}$) of the upper continuum of the Dirac levels. In the same time, the $a-$ particle energy, $-\varepsilon_{p}$, is negative, and is below the upper boundary ($-m_{a}$) of the lower continuum of the Dirac levels.
In this region, the $b-$ particle momentum $q\geq 2.659\sqrt{m_{b}m_{a}}$. We can estimate the characteristic radius of its motion in the BIC state: $R_{eff} < (2.659\sqrt{m_{b}m_{a}})^{-1}$. As for the 
$a-$ particle, there are no restrictions on its momentum. Moreover, the $a-$ particle momentum increases sharply with $q$. It is important that this increase in the momenta $p$ and $q$ does not lead to a change in the energy $E=m_{b}+m_{a}+{\mathcal E}$. 

\par Note that for normal bound states, the finding a particle in lower continuum states is not uncommon. For example, in the bound states of the hydrogen, the electron is also characterized by a wave function in the lower continuum. But the probability of being in them is small, of the order of $\simeq \alpha^2$.

\par In the second resonant region:
\begin{eqnarray}         
\psi({\bf p},{\bf q};E)&=&-\frac{\alpha}{8\pi^2\omega_{q}\varepsilon_{p}}\frac{\Lambda_{a}^{-}\Lambda_{b}^{+}}{E+\varepsilon_{p}-\omega_{q}+i\delta}
\nonumber\\& &
\int \frac{d{\bf k}}{k^2}(1-{\bm \alpha}_{a}{\bm \alpha}_{b})\psi({\bf p}+{\bf k};{\bf q}-{\bf k}).
\label{eq:15}
\end{eqnarray}

\par The solution of the integral Eq.~(\ref{eq:15}) is sought in the form:
\begin{eqnarray}         
\psi({\bf p},{\bf q};E)&=&\sqrt{\frac{\delta}{\pi}}
\frac{\xi({\bf p},{\bf q};E)}{E+\varepsilon_{p}-\omega_{q}+i\delta}.
\label{eq:16}
\end{eqnarray}

From Eq.~(\ref{eq:16}) we obtain the two-particle probability density:
\begin{eqnarray}         
\vert \psi({\bf p},{\bf q};E) \vert^{2}&=&\vert \xi({\bf p},{\bf q};E) \vert^{2} \delta(E+\varepsilon_{p}-\omega_{q}).
\label{eq:17}
\end{eqnarray}
with 
\begin{eqnarray}         
\int \int d{\bf p}d{\bf q}\vert \xi({\bf p},{\bf q};E) \vert^{2} \delta(E+\varepsilon_{p}-\omega_{q})&=&1
\label{eq:18}
\end{eqnarray}

\par One can see that the two-particle probability density Eq.~(\ref{eq:17})  is defined only on the resonant region presented in Fig. 1. Nevertheless, the function $\psi$ can be defined in the whole momentum space. Substituting Eq.~(\ref{eq:16}) into Eq.~(\ref{eq:15}), we obtain:
\begin{eqnarray}         
\xi({\bf p},{\bf q};E)&=&\frac{i\alpha}{2\pi}\frac{\Lambda_{a}^{-}\Lambda_{b}^{+}}{4\omega_{q}\varepsilon_{p}}              
\int_{\{{\bf p},{\bf q}\}\in D_{2}} \frac{d{\bf k}}{({\bf p}-{\bf k})^2}(1-{\bm \alpha}_{a}{\bm \alpha}_{b})
\nonumber\\& &
\psi({\bf k};{\bf q}+{\bf p}-{\bf k}) \delta(E+\varepsilon_{{\bf k}}-\omega_{{\bf q}+{\bf p}-{\bf k}}).
\label{eq:19}
\end{eqnarray}
Here $D_{2}$ is denoted the second resonant region. 

\par Eq.~(\ref{eq:19}) is the integral equation with the pure imaginary kernel. As a result, $\xi({\bf p},{\bf q};E)$ must be the complex function. Taking into account that
the particle $b$ is non-relativistic and the relativistic particle $a$ is described by bispinor wave function, Eq.~(\ref{eq:15}) is the four  
interrelated integral equations for the functions of three variables, namely, $p$, $q$ and the angle between these momenta. The angle presents the correlations in the particles motion in the BIC state.  

\section{The effective interaction constant}

In Eq.~(\ref{eq:19}) the energy $E=m_{b}+m_{a}+\gamma m_{a}$, the integration over $\bf k$ is carried out in the second resonant region $D_{k}$ area, which for Eq.~(\ref{eq:19}) is defined as:
\begin{eqnarray}         
m_{b}+(\gamma+1)m_{a}+\sqrt{m_{a}^2+p^2}=\sqrt{m_{b}^2+({\bf q}+{\bf p}-{\bf k})^2}.
\label{eq:20}
\end{eqnarray}

\par From Eq.~(\ref{eq:20}) we conclude that in the second resonant region the $a-$ particle is relativistic and its momentum $p>m_{a}$. Taking into account $m_{b}>>m_{a}$, the $b-$ particle  is non-relativistic.  Its energy can be written as $\omega_{q}=m_{b}+\frac{q^2}{2m_{b}}$, and its momentum $q\propto \sqrt{m_{a}m_{b}}$. Then Eq.~(\ref{eq:7}) is reduced to:   
\begin{eqnarray}         
\Lambda^{+}_{b} \to 2m_{b}\left(\begin{array}{cc} {1} & 0\\0 & {0} \end{array}\right)_{b}.
\label{eq:21}
\end{eqnarray}
Also, the interaction of the particles through the vector potential can be omitted since  
\begin{eqnarray}          
{\bm \alpha}_{a}{\bm \alpha}_{b}\simeq \sqrt{\frac{m_{a}}{m_{b}}}<<1.
\label{eq:22}
\end{eqnarray}

\par Due to the symmetry of the problem, the functions $\xi$ can depend on absolute values of the vectors p and q, and the angle between them $\theta$, that is $\xi(p,q;\theta)$.  
Integrating over ${\bf k}$ in Eq.~(\ref{eq:19}), we can, without loss of generality, assume that ${\bf q}+{\bf p}$ is directed along the $z-$ axis.
Then Eq.~(\ref{eq:19}) is  reduced to: 
\begin{displaymath}
\xi(p,q;\theta)=i\alpha\frac{m_{b}}{\vert {\bf q}+{\bf p} \vert} \frac{\varepsilon_{{\bf p}}- {\bm \alpha}_{a}{\bf p}- m_{a}\beta_{a}}{2m_{a}} 
\end{displaymath}
\begin{displaymath}
\int_{D_{k}} \frac{kdk}{\sqrt{(k^2+p^2-2kp\cos \theta_{k} \cos \theta_{a})^2-4k^2 p^2 \sin^2 \theta_{k} \sin^2 \theta_{a}}}
\end{displaymath}
\begin{eqnarray}          
\xi(k,\vert {\bf q}+{\bf p}-{\bf k} \vert;\theta_{*}). 
\label{eq:23}
\end{eqnarray}
Here $\theta_{a}$ is the polar angle of ${\bf p}$, $\theta_{k}$ is the polar angle of ${\bf k}$, $\theta_{*}$ is the angle between ${\bf q}+{\bf p}-{\bf k}$ and ${\bf k}$.

\par In Eq.~(\ref{eq:23}) $p\propto m_{a}$ and $k\propto m_{a}$. With the exception of $\xi$, the expression under the integral on the right side of ~(\ref{eq:23}) is dimensionless.
Taking into account $q\propto \sqrt{m_{a}m_{b}}$ and $q>>p$, the factor before the integral is:
\begin{eqnarray}          
\frac{m_{b}}{\vert {\bf q}+{\bf p} \vert} \frac{\varepsilon_{{\bf p}}- {\bm \alpha}_{a}{\bf p}- m_{a}\beta_{a}}{2m_{a}}
\propto \sqrt{\frac{m_{b}}{m_{a}}} 
\label{eq:24}
\end{eqnarray}

Using $m_{b}= 1836 m_{a}$, from Eq.~(\ref{eq:24}) we obtain:
\begin{eqnarray}          
\alpha_{eff}=\alpha\sqrt{\frac{m_{p}}{m_{e}}}=0.313.
\label{eq:25}
\end{eqnarray}

This value $\alpha_{eff}$ can be regarded as the effective interaction constant. The fact that $\alpha_{eff}>>\alpha$ is caused by the resonance of the electromagnetic interaction between particles in the BIC state.

\section{Conclusion}

\par In present work, the theory of BIC states was supplemented with the conception of the resonance of interparticle interaction.  Using the Bethe-Salpeter equation, the resonant regions in momentum space are found in which there is the sharp increase in the electromagnetic interaction between the two particles. These resonant regions, together with correlations in particles motion, determine the confinement mechanism of the composite particle  in the BIC state. Due to the interaction resonance, the positive binding energy can be of the order of the mass of the lighter particle from this pair.

\par The composite particle from the two fermions is a boson with the integer spin, 0 or 1. However, it should be taken into account that $m_{b}>>m_{a}$. In a real situation, this will lead to the fact that the magnetic moment of particle $a$ is much greater than the magnetic moment of particle $b$. Then, in many experiments, which are considered as direct ones to determine the spin, this composite particle would represent the spin equal to 1/2. For example, \cite{bib10,bib11} used an experimental setup similar to that of Stern and Gerlach. In this experiment, the particle spin is determined from the splitting of a particle beam when it is passed through a highly inhomogeneous magnetic field.  There is no doubt that the splitting of the beam of these composite bosons into two components would certainly be observed. However, the spin of this composite particle is not equal to 1/2. Note that a beam of hydrogen atoms in the ground state would also split into two components, despite the fact that this atom is the boson. As is known, in such experiment, the intrinsic magnetic moment of the electron was established \cite{bib12}. 
Other experiments are also known \cite{bib13,bib14,bib15}, the data of which, for the above reason, do not answer the question what is the spin of the composite particle?

\par Finally, the composite particle in the BIC state exists only in the free state. In general, the physical properties of this particle should be studied only when it is free.
It is obvious that in the other composites, the composite particle loses its individuality.

\end{document}